\documentclass[a4paper,11pt]{article}
\usepackage{pos}

\title{Detecting Magnetar Giant Flares with the Moon Burst Energetics All-sky Monitor (MoonBEAM)}
\ShortTitle{Detecting Magnetar Giant Flares with MoonBEAM}

\author*[a]{O.J. Roberts}
\author[b]{E. Burns}
\author[a]{A. Goldstein}
\author[c]{C.M. Hui}

\onbehalf{for the MoonBEAM Team}

\affiliation[a]{Universities Space Research Association, 320 Sparkman Dr., Huntsville, AL 35805, USA}

\affiliation[b]{Department of Physics \& Astronomy, Louisiana State University, Baton Rouge, LA 70803, USA}

\affiliation[c]{Astrophysics Branch, ST12, NASA Marshall Space Flight Center, Huntsville, AL 35812, USA}

\emailAdd{oroberts@usra.edu}

\abstract{Magnetars are slowly-rotating neutron stars with extremely strong magnetic fields (10$^{13-15}$~G) that rarely produce extremely bright, energetic giant flares. Magnetar Giant Flares (MGFs) begin with a short (200 ms) intense flash, followed by fainter emission lasting several minutes that is modulated by the magnetar spin period (typically 2-12 s). Over the last 40 years, only three MGFs have been observed within our Galaxy and the Magellanic Clouds, which all suffered from instrumental saturation due to their extreme intensity. It has been proposed, that extragalactic MGFs masquerade as a small subset of short Gamma-ray Bursts (GRBs), noting that the sensitivity of current instrumentation prevents us from detecting the pulsating tail to distances slightly beyond the Magellanic Clouds. However, their initial bright flash is readily observable out to distances of < 25 Mpc. In this presentation, we will evaluate the spectral and temporal behaviors of MGFs using recent observations from events such as GRB200415A, to differentiate them from other progenitors, such as  short GRBs. We then present an overview of the Moon Burst Energetics All-sky Monitor (MoonBEAM), which will attempt to discover more of these events, providing highly sensitive data that will help unravel the nature of these phenomena further in an attempt to better understand their emission mechanisms comparatively with GRBs. In doing so, MoonBEAM will help provide a comprehensive picture of energetic astrophysical phenomena, a key goal of the Astro2020 decadal survey.}

\ConferenceLogo{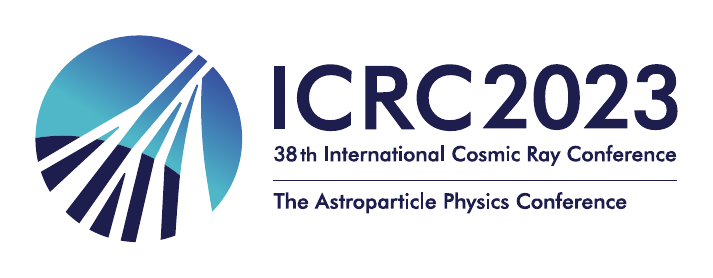}
\FullConference{%
38th International Cosmic Ray Conference (ICRC2023)\\
  26 July - 3 August, 2023\\
  Nagoya, Japan}

\begin{document}
\maketitle

\section{Introduction} \label{sec:intro}

Neutron stars are produced from the core-collapse of massive stars (i.e., $\sim$20~$M_{\odot}$), providing an opportunity to test the fundamental laws of physics due to the extreme nature and environment of the resulting system. ``Magnetars" are highly magnetized ($10^{13-15}$\,G~\cite{ThompsonDuncan1995}), isolated, slow-rotating neutron stars that episodically emit luminous, short ($\sim$10$^{-2}$ s to $\sim$1 s), energetic (typically 10$^{36-41}$~erg) X-ray bursts. During ``outbursts", they emit longer (1--50~s), more energetic (10$^{41-43}$ erg) ``intermediate flares". Rarely, they emit giant flares, extremely bright, energetic (10$^{44-47}$ erg) events that are characterized by an initially short ($\sim0.1-0.2$\,s) intense flash during its onset that gradually fades to a fainter, longer-lasting tail of emission (up to hundreds of seconds long), that oscillates at the spin frequency of the neutron star and provides the ``smoking gun" evidence of the origin of the emission~\citep{Kaspi2017}.

The last 40--50 years of almost continuous monitoring of the X-ray and gamma-ray transient sky, has resulted in only three MGFs being observed in our local galactic group: GRB\,790305B in 1979 (from the Large Magellanic Cloud~\citep{Mazets1979}), GRB\,980827 from SGR 1900+14~\citep{Hurley1999} and GRB\,041227 from SGR 1806-20~\citep{Palmer2005} (both from the Milky Way). The data from these observations suffered heavily from debilitating instrumental ``saturation" effects, such as dead time and pulse pileup, that prevented clean observations of the first second of each event. Consequently, a true snapshot of the onset of MGFs from local events has remained elusive. However, the high peak luminosity of the relatively short ($\sim$10$^{-1}$s) initial MGF ``spike" means that it is readily observable out to distances of $\sim 10$ Mpc~\citep{Burns2021}. Consequently, it's been proposed that extragalactic MGFs make up $\sim2\%$ of the short Gamma-ray Burst (sGRB) population~\citep{Burns2021}. The limited sensitivity of current instrumentation prevents the detection of the pulsating tail at distances greater than the Magellanic clouds. Due to their transient nature, the best chance of confirming an extragalactic MGF candidate is by using sensitive instruments to instantaneously monitor as much of the unocculted gamma-ray sky as possible, using the arrival time of photons to widely scattered satellites in an InterPlanetary Network (IPN) to triangulate the transient emission to regions of the order of several arcminutes or degrees, that overlap with nearby galaxies. Five extragalactic MGF candidates were found this way: GRB\,051103~\citep{Ofek2006}, GRB\,070201~\citep{Ofek2008}, GRB\,070222~\citep{Burns2021}, GRB\,180128A~\citep{Trigg2023}, GRB\,200415A~\citep{Roberts2021}, each triangulated to a nearby star-forming galaxy (M81, M31, M83 and NGC 253 (twice), respectively). Unlike the three galactic MGFs, the emission from these five MGFs over several Mpc means the recorded signal is largely unaffected by instrumental saturation effects.

While only eight MGF candidates have ever been recorded, it is believed that the intrinsic volumetric rate of MGFs is significantly higher than previous thought, with $R_{MGF} =3.8^{+4.0}_{-3.1}\times10^{5}~\mathrm{Gpc^{-3}~yr^{-1}}$~\citep{Burns2021}. It is believed the high volumetric rate implies a common progenitor resulting from a relatively common astrophysical event (e.g., core-collapse supernovae and other end-of-life events). While more controversial, such a high rate could also imply that magnetars are capable of producing multiple giant flares during their relatively short lifetimes, implying that GRBs may repeat~\citep{PopovStern2006}. 

Motivated by these outstanding scientific queries, we present The Moon Burst Energetics All-Sky Monitor (MoonBEAM), a highly sensitive gamma-ray monitor that will seek to better understand the onset of MGFs. In this proceeding, we discuss the prototypical MGF; GRB 200415A, how we can use the temporal and spectroscopic information garnered from this study to elaborate on a progenitor origin, and how MoonBEAM will use this to answer its science objectives on MGFs.

\section{GRB 200415A, the prototypical extragalactic MGF}

The best example of an extragalactic MGF occurred on the 15$^{th}$ of April 2020, when GBM was triggered by an extremely short, bright, energetic event called GRB 200415A~\citep{Bissaldi2020}. The event was detected by several other instruments~\citep{Svinkin2021,Omodei2021,ACJT2021}, and was found with the Swift-BAT Gamma-ray Urgent Archiver for Novel Opportunities (GUANO)~\cite{Tohuvavohu2020} pipeline. GRB 200415A was triangulated by the IPN to a 17 square arcminute region centered at RA=11.88$^{\circ}$ (00h 47m 32s), Dec.=-25.263$^{\circ}$ (-25d 15' 46", J2000), which spatially overlaps with the Sculptor Galaxy (NGC 253), an active star-bursting spiral galaxy $\sim$3.5 Mpc away~\citep{Svinkin2021}, with a chance coincidence of association of 1 in 230,000~\citep{Burns2021}. 

\begin{figure}
    \centering
    \includegraphics[width=\linewidth, trim={0 2cm 0 1cm},clip]{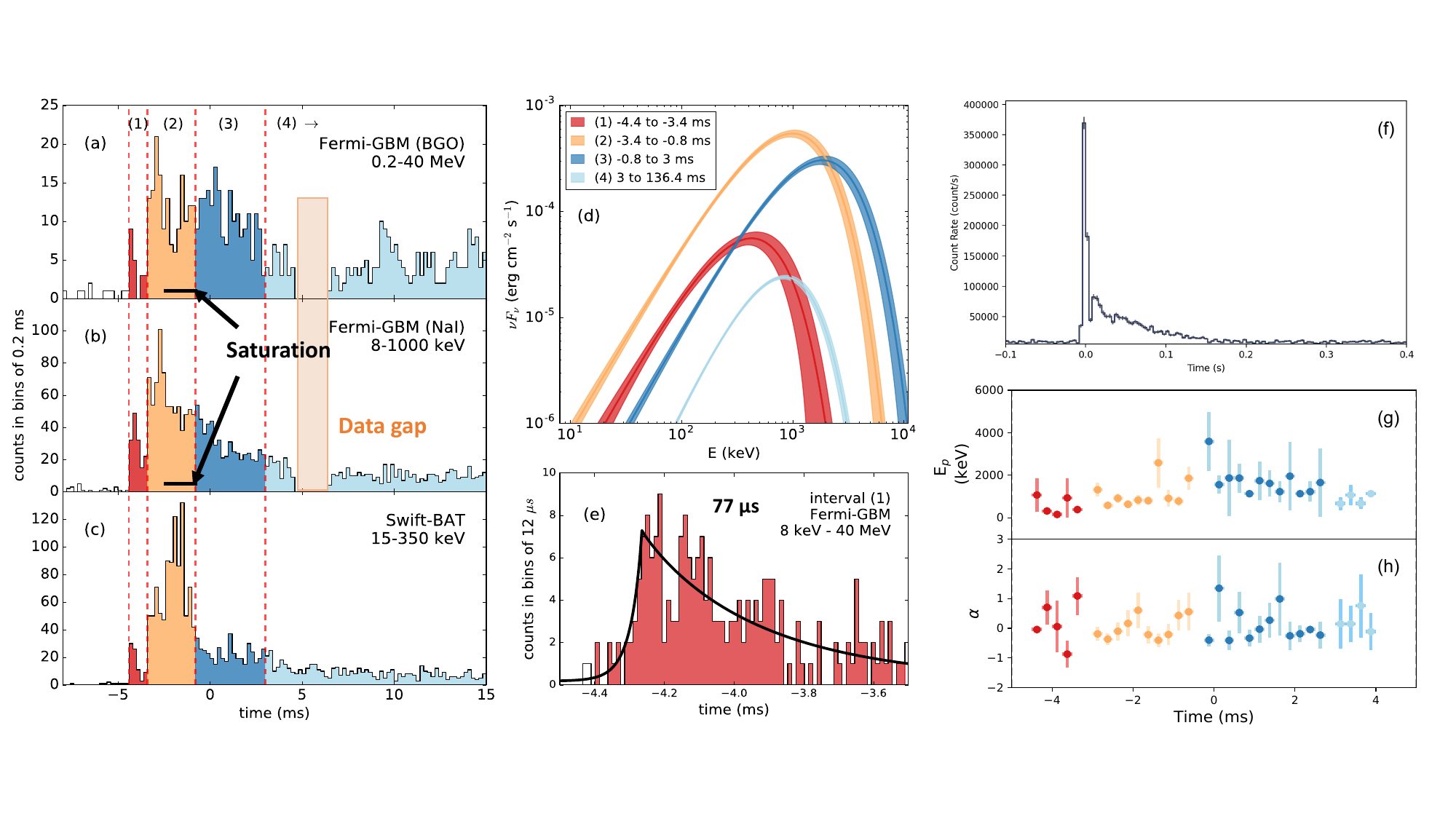}
    \caption{\textbf{Left Panel:} A 25~ms interval, binned to 200~$\mu$s, showing the first 20~ms of GRB 200415A, with the data gap and saturated parts of the GBM lightcurves (a and b) shown. The Swift-BAT signal that was used to recover the GBM spectral information is shown in sub-panel (c). The four intervals are color-coded according to the legend in the middle panel. \textbf{Middle Panels:} The $\nu$F$_{\nu}$ spectral density plot showing the energetic behavior of the MGF (d). A fit to the risetime of the event, found to be about 77~$\mu$s (e). \textbf{Right Panels:} Top, GBM lightcurve binned to 4~ms (f). Bottom, the sub-ms (250~$\mu$s) variation over a 10~ms interval in E$_{p}$ (g) and $\alpha$ (h). Figure is adapted from \citep{Roberts2021}.}
    \label{fig:200415a}
\end{figure}

The temporal and spectral behavior of the peak of GRB 200415A is shown in Fig.~\ref{fig:200415a}. GRB 200415A was found to have a T$_{90}$ duration of 140.8$^{+0.5}_{-0.6}$~ms (1$\sigma$). The rise time (10 to 90\%) was determined to be $T_{\rm rise}$=77$\pm$23 $\mu$s (1$\sigma$)~\citep{Roberts2021}. Additional timing analysis carried out found a broad quasi-periodic oscillation candidate (a possible signature of seismic vibrations), at a frequency of $\nu$$\sim$180~Hz at $\sim 2.5\sigma$ significance~\citep{Roberts2021}, with other higher-frequency candidates reported by ASIM~\citep{ACJT2021}.

Time-integrated and time-resolved spectral analysis of several intervals of the GBM data were performed, fitting multiple spectral models. A power-law with an exponential cutoff (COMPT) model was found to fit the time-integrated data the best, as well as the four intervals (red, orange, blue) shown in Fig.~\ref{fig:200415a}. The spectral parameters are shown in Table~\ref{tab:1} (adapted from \cite{Roberts2021}). The GRB 200415A event rate from T$_{0}$-2.4 ms to -0.8 ms exceeded the instrument bandwidth rate (375~kHz for all detectors~\citep{Meegan2009}) in the GBM time-tagged event data. Additionally, there is a $\sim$1.8~ms data gap from T$_{0}$+4.6~ms to +6.4~ms. The highest energy photon detected by GBM was $\sim$ 3~MeV. The time-integrated isotropic equivalent energy output was found to be $E_{\rm iso}=(1.5\pm0.02) \times 10^{46}$ erg, and the peak isotropic luminosity, $L_{\rm iso,\max}= (1.5 \pm0.1) \times 10^{48} $~ erg s$^{-1}$. The total event luminosity was found to be $L_{\rm iso}= (1.1 \pm0.2) \times 10^{47} $~ erg s$^{-1}$, after correcting for the portion of the spectrum where the bandwidth was exceeded with the Swift-BAT data. Fig.~\ref{fig:200415a} also shows sub-ms variations of $E_{\rm p}$ over the first 10\,ms of the event. The photon index stays relatively constant/flat over the whole event ($\alpha \sim 0$). The GBM data also revealed differing exponential decay trends in both energy flux (${\cal F}$) and $E_{\rm p}$ of $\tau$ = 45 $\pm$ 3~ms and $\tau$ = 100 $\pm$ 1~ms, respectively, over the majority of the MGF (interval 4)~\citep{Roberts2021}. This behavior has been observed in other extra-galactic GF candidates~\citep{Svinkin2021}. A strong ${\cal F} \propto E_{\rm p}^2$ correlation was also found~\citep{Roberts2021}. 

\begin{table}[]
    \centering
    \resizebox{0.7\columnwidth}{!}{%
    \begin{tabular}{|c|c|c|c|c|c|}
    \hline
Time  & $ E_{\rm peak}$ & $\alpha$ & Flux & $L_{\gamma, iso}$& E$_{\gamma, iso}$ \\
		(ms) & (keV) &  & ($10^{-5}$ erg cm$^{-2}$ s$^{-1})$ &   ($10^{47}$ erg s$^{-1}$)& ($10^{45}$ erg)  \\ \hline
(1) $-4.4$ to $-3.4$   &$428\pm71 $  &$-0.1 \pm 0.2$ & $9.9\pm1.2$          &$ 1.5 \pm  0.2$ & $0.2\pm0.1$ \\
(2) $-3.4$ to $-0.8$   &$997\pm77 $  &$-0.2\pm0.1 $ & $33.7\pm1.5$       &$15.3  \pm  1.3 $ &$ 4.0 \pm  0.3$ \\
(3) $-0.8$ to $3.0$     &$1856\pm155$ &$-0.1 \pm 0.1$ & $17.5\pm 0.8$      &$ 8.3 \pm  0.4$ &$ 3.2 \pm  0.2$  \\
(4)	$3.0$ to $136.4$   &$846\pm39$   &$ 0.3 \pm 0.1$ & $2.7\pm 0.1$    &$ 0.6 \pm  0.1$&$ 7.8 \pm  0.4$  \\
	T$_{90}$ (140.8) &  &       & & $1.1 \pm 0.2$ & $\mathbf{ 15.1 \pm 2.5}$ \\
 \hline
    \end{tabular}
    }
    \caption{The spectral parameters, luminosity and emitted energy for COMPT model fits to the four time intervals identified in Fig.~\ref{fig:200415a}, relative to the GBM trigger time. All errors are at the 1$\sigma$ confidence level. \label{tab:Table1}}
    \label{tab:1}
\end{table}

\section{Temporal and Spectral Characteristics of MGFs}\label{sec:temporalspecchar}

The temporal and spectral analysis of GRB 200415A provides a remarkably clear snapshot into the nature of the possible progenitor, an energy release or MGF due to a crustal fracture (or ``starquake") of a magnetar, triggered by hot plasma being deposited into the inner magnetosphere from sub-surface fields. This picture can be constructed from the spectroscopic and temporal results in multiple ways, addressed in each of the following subsections.

\subsection{The ``Compton Cloud": L$_{iso}$ and $\alpha$}
The L$_{iso}$ implies a high electron density ($\geq$10$^{29}$~cm$^{-3}$) when the plasma is near the stellar surface, initially, highly opaque to electron scattering ($\lambda_{s}$ $\leq$ 10$^{-5}$~cm) out to altitudes of roughly R $\geq$ 10$^{9-10}$~cm. The photon index ($\alpha \sim -0.2 \longrightarrow 0.3$) results from high n$_{e}$ densities seeding Comptonization that results in a spectrum approaching a modified blackbody (Wien). The source at Sculptor's distance must be a compact, solar-mass-scale neutron star or black hole in order to reproduce an emission region as optically thick as GRB 200415A. This behavior in $\alpha$ is inconsistent with traditional synchrotron emission scenarios involved for explaining GRB spectra. 

\subsection{Relativistic outflow: E$_{p}$, ${\cal F}$, L$_{iso}$, MeV-band emission}

A highly relativistic outflow ($>$0.98c) is implied from L$_{iso} >> L_{edd}$ (10$^{47-48}$ $>>$ 10$^{38}$ erg s$^{-1}$) and the MeV-band emission. The emission above 511-keV from $\gamma\gamma \longrightarrow e^{+}e^{-}$ interactions (QED pair production) were used to provide a lower bound to the wind bulk Lorentz factor, $\Gamma$. Wind transparency to 3 MeV photons suggests a $\Gamma > E_{max}/m_{e}c^{2}$ $\sim$ 6, and provides more restrictive values on previous measurements of E$_{max}$ from the SGR 1900+14 and SGR 1806-20 giant flares. The Fermi-LAT detection of several photons (ranging from 480 MeV to 1.7 GeV), from 19 seconds after the emission detected by GBM and BAT, implies $\Gamma$ is much higher~\citep{Omodei2021}. The photon energy flux is considerably less than what is expected for a Plank distribution from a relativistic wind that yields similar E$_{p}$ values in the observer frame, which strongly contrasts the GRB fireball scenario. The exponentially decaying tail of the flux after the initial spike ($\tau \sim 45$ms), for which there is a distinctive ${\cal F}\propto E_{\rm p}^2$ correlation~\citep{Roberts2021} is the hallmark of a relativistic wind, and corroborates the conjecture that the outflow is highly relativistic. 

\subsection{Radiation Lighthouse Model: E$_{p}$, ${\cal F}$, $\nu$F$_{\nu}$}

The sub-millisecond spectral evolution and behavior in $\nu$F$_{\nu}$, shown in Fig.~\ref{fig:200415a} shows a relativistic lighthouse beaming effect. As this Doppler-boosted cone sweeps across our line-of-sight, the intensity increases, peaks and then declines, accompanied by spectral hardening and softening. This is a convolution of influences of stellar rotation and radiation collimation (cone subtending angle 1/$\Gamma$). When coupling an average neutron star spin period (P=8 s) and a $\Gamma$$\sim$30 bulk outflow into the equation $\tau$ $\sim$ P/2$\pi$$\Gamma$ for a rotating beam, the 45~ms decay time of the ``tail" of GRB 200415A, is reproduced. This corresponds to $\Delta\Theta$$\sim$2$^{\circ}$ rotation of the star. 

\subsection{Risetime and differences with GRB catalogs}

Comparing GRB 200415A with sGRBs from the GBM catalog~\citep{AvK2020}, GRB 200415A was found to be in the 97.5th percentile of the sGRB distribution when comparing the 64~ms peak photon flux with sGRBs. Similarly, E$_{p}$ was found to be in the 79$^{th}$ percentile and $\alpha$ in the 89$^{th}$ percentile, where the latter was also found to lie on the outside of the $\alpha$ distribution for sGRBs from the BATSE burst catalog~\citep{Kaneko2006}. The 77~$\mu$s risetime (shown in Fig.~\ref{fig:200415a}, is characteristic of a GF onset and is much shorter for any event in GBM or BATSE catalogs, even those with the shortest variability (i.e., 100 $\mu$s for GRB 910711~\citep{Bhat1992}).

\section{The Moon Burst Energetics All-Sky Monitor (MoonBEAM)}

Summarizing, we have identified a lot of the temporal and spectral observations from ``GRB" 200415A as markedly different from classic GRB scenarios and are more suggestive of a MGF from an extragalactic magnetar. The spectroscopic and temporal tools developed from this study can be used to unmask sGRBs, providing more details of the onset of MGFs, devoid of instrumental saturation. This gives new insight into the long elusive mechanism and provides more tools to classify extragalactic MGFs, other than identification of a host galaxy. Such developments enable the active pursuit of similar events as fundamental science objectives for future missions, like MoonBEAM.  

The Moon Burst Energetics All-sky Monitor (MoonBEAM) is a proposed gamma-ray instrument intended for cis-lunar orbit with highly sensitive detectors to observe relativistic astrophysical explosions in near-real time over the entire sky instantaneously for 2.5 years~\citep{Fletcher2023}. The mission objective to evaluate relativistic transients produced from the mergers of compact objects, core-collapse supernovae, or MGFs, will be achieved by characterizing GRB progenitors and their multi-messenger/-wavelength signals to identify the necessary conditions to launch an astrophysical jet and determine the mechanism of observed high-energy emission within a relativistic outflow. 

MoonBEAM's orbit will mean an almost instantaneous view of the whole sky as there will be $<<1\%$ occultation by the Earth and the Moon at closest approach and a high duty cycle of $\geq$ 94\%. 
MoonBEAM will be able to localize events in a similar way to GBM~\citep{Meegan2009} and will be a valuable instrument as part of the IPN, as photons will be detected from astrophysical events with a difference of up to 2 light seconds relative to other satellites in low-earth orbit, providing refined IPN localizations of transient events due to longer baselines between independent satellites. Such refined localizations will solve long-standing issues with localizing transients such as extragalactic MGFs and GRBs without a pointing instrument such as a telescope, which usually has a very limited field of view (FoV).

MoonBEAM will address its scientific objectives by using its suite of six phosphor sandwich or ``Phoswich" scintillation detectors. Each phoswich detector comprises a thick (3.2~cm) scintillator layer of sodium-doped Caesium Iodide (CsI:Na), optically coupled together with a thinner (1.5~cm) thallium-doped sodium iodide (NaI:Tl) scintillator. The scintillators each have an active diameter of 14~cm and are coupled to Hamamatsu flat panel photomultiplier tubes. The response of the thin NaI:Tl detectors changes rapidly as the source angle deviates from normal and is used for localization, while the thick CsI:Na crystals maintain effective area over a wider FOV due to the contribution of their broader sides, improving sky coverage and high-energy sensitivity. The six detectors are positioned such that the full sky is continuously monitored with a near-uniform sensitivity, effectively detecting gamma-ray photons over an energy range of 10--5000~keV. MoonBEAM will acquire continuous data, which is binned as 16 pseudo-logarithmically spaced energy bins, and is binned temporally to 64~ms. The triggered data is binned as 32 pseudo-logarithmically spaced energy bins with a temporal resolution of 30~$\mu$s. The detector timing accuracy is 10~ms. 

 
Based on the local rate ($<$ 25~Mpc) of MGFs~\citep{Burns2021}, it is estimated that MoonBEAM will make 50 observations of MGFs during its lifetime due to the instrument capabilities and sensitivity (trigger sensitivity at 64~ms of 3.0~$ph~s^{-1}~cm^{-2}$), with a limiting flux of $2.2 \times 10^{-7}~erg~cm^{2}~s^{-1}$ (50-300~keV, 1~s, limiting luminosity of $10^{46}~erg~s^{-1}$). It also assumes a FoV of $>$99\%, a duty cycle of $\geq$94\%. It is estimated that at least two MGFs will be positively identified over the mission lifetime. 

MGF candidates detected by MoonBEAM will be confirmed by using the IPN to identify a local host galaxy. Preliminary work based on GBM GRBs with IPN localizations has found that MoonBEAM will need to localize to $\sim$10~deg$^{2}$ in order to classify an MGF at the 90\% confidence level and 100~deg$^{2}$ to classify sufficiently bright MGFs at the 75\% confidence level. As an instrument of the IPN, the median width of timing annuli for triggered events should be of the order of several degrees~\citep{Fletcher2023}. In instances where the IPN cannot localize an MGF to a host galaxy, temporal and spectral analysis as discussed in Section~\ref{sec:temporalspecchar}, will be used. The 30 $\mu$s temporal resolution of the data and the energy range of the detectors will ensure that a lot of the key signatures detected by GBM will be captured by MoonBEAM, providing the necessary tools to identify extragalactic MGFs masquerading as sGRBs. 

MoonBEAM will answer key outstanding questions about whether individual magnetars produce multiple giant flares~\citep{PopovStern2006}, or whether magnetars undergo a phase transition culminating in a single giant flare event that substantially depletes the energy reservoir of the system, preventing further giant flares~\citep{Ioka2001}. To prove MGFs happen multiple times, the rate of MGFs needs to be better constrained and shown to exceed the rate of core-collapse supernovae, believed to be one of the main neutron star formation channels. A confirmation of at least two MGFs during the MoonBEAM mission lifetime of 2.5 years, combined with a historic sample~\citep{Burns2021}, would be sufficient to resolve these questions. 

\section{Summary}

The onset of MGFs within our galaxy saturate instrumentation, preventing information on its initiation mechanism. At extragalactic distances, several MGF candidates have been identified, with the most notable being GRB 200415A. The IPN localization of this event and the associated temporal and spectral information from GBM and BAT observations allowed a rare, in-depth study of the brightest, most energetic portion of an MGF. Specifically, the observation of GRB 200415A created new temporal and spectroscopic traits to search for when unmasking sGRBs in archival data, and enables the search for similar future events. Leveraging what was learned from GRB 200415A, MoonBEAM will answer whether individual magnetars produce multiple giant flares, as well as provide new insight into the physical mechanism for their generation as part of multiple scientific objectives to understand the most exotic and extreme relativistic transients in the observable Universe.

\end{document}